\documentclass[conference]{IEEEtran}
\IEEEoverridecommandlockouts
\usepackage{cite}
\usepackage{amsmath,amssymb,amsfonts}
\usepackage{graphicx}
\usepackage{textcomp}
\usepackage{xcolor}
\usepackage[T1]{fontenc}
\usepackage{hyperref} % Links
\usepackage[utf8]{inputenc}
\usepackage{xspace}
\usepackage{subfig}
\usepackage{longtable}
\usepackage{multirow}
\usepackage{cleveref}
\usepackage{url}
\usepackage{soul}
\usepackage{wrapfig,booktabs}
\usepackage{algorithm}
\usepackage{float} 
\usepackage[noend]{algpseudocode}
\usepackage{xspace}
\usepackage[inline]{enumitem}
\usepackage[normalem]{ulem} 

\newcommand{\oursystem}{ThreatLinker\xspace}

\def\BibTeX{{\rm B\kern-.05em{\sc i\kern-.025em b}\kern-.08em
    T\kern-.1667em\lower.7ex\hbox{E}\kern-.125emX}}
    
\begin{document}

%%%%% USAGE OF REVIEW COMMAND: both parameters are optional
%% \review{text to remove}{text to add}
\newif\ifreview
\reviewfalse % \reviewtrue (to activate) or \reviewfalse (to disable) review mode
% Base annotation macro
\ifreview
    \newcommand{\ygg}[2]{\fbox{\bfseries\sffamily\scriptsize#1}{\sf\small$\blacktriangleright$\textit{#2}$\blacktriangleleft$}}
    \newcommand{\annote}[2]{\ygg{\textsc{#1}}{#2}}
    \newcommand{\review}[2]{\annote{}{\textcolor{red}{\sout{#1}}~\textcolor{blue}{#2}}}
\else
    \newcommand{\review}[2]{{#2}}
\fi

\title{\oursystem: An NLP-based Methodology to Automatically Estimate CVE Relevance for CAPEC Attack Patterns
}

\author{\IEEEauthorblockN{Andrea Ciavotta}
\IEEEauthorblockA{\textit{Sapienza University of Rome} \\
Rome, Italy \\
{\small ciavotta.1491111@studenti.uniroma1.it}}
\and
\IEEEauthorblockN{Alessandro Palma}
\IEEEauthorblockA{\textit{Sapienza University of Rome} \\
Rome, Italy \\
{\small palma@diag.uniroma1.it}}
\and
\IEEEauthorblockN{Simone Lenti}
\IEEEauthorblockA{\textit{Sapienza University of Rome} \\
Rome, Italy \\
{\small lenti@diag.uniroma1.it}}
\and
\IEEEauthorblockN{Silvia Bonomi}
\IEEEauthorblockA{\textit{Sapienza University of Rome} \\
Rome, Italy \\
{\small bonomi@diag.uniroma1.it}}
}

\maketitle

\begin{abstract}
Threat analysis is continuously growing in importance due to the always-increasing complexity and frequency of cyber attacks. Analyzing threats demands significant effort from security experts: different cybersecurity knowledge bases support this task, but manual efforts are required to correlate heterogeneous sources into a unified view that would enable a more comprehensive assessment.
To address this gap, we propose ThreatLinker, a methodology leveraging Natural Language Processing (NLP) to effectively and efficiently associate Common Vulnerabilities and Exposure (CVE) vulnerabilities with Common Attack Pattern Enumeration and Classification (CAPEC) attack patterns.
The proposed technique combines semantic similarity with keyword analysis to improve the accuracy of association estimations.
We contributed a larger dataset for CVE-CAPEC correlation, and experimental evaluations demonstrate superior performance compared to state-of-the-art models.
\end{abstract}


\section{Introduction}
\label{sec:introduction}

The growing complexity of cyberattacks necessitates innovative approaches to threat analysis and risk mitigation.
Various cybersecurity knowledge bases have been developed to support these efforts, offering structured and comprehensive information on vulnerabilities, attack techniques, and corresponding mitigation strategies~\cite{chi2023prioritizing}.
%Since they focus on different aspects, it is crucial to integrate these knowledge bases to create a unified, comprehensive view of the problem at hand for improved threat analysis.
Although these resources often focus on different facets of the cybersecurity landscape, integrating them is essential to provide a unified and holistic view that enhances threat analysis capabilities~\cite{vermeer2023alert}.
Among the most prominent of these knowledge bases, the Common Vulnerabilities and Exposures (CVE) catalog contains detailed records of publicly disclosed vulnerabilities, while the Common Attack Pattern Enumeration and Classification (CAPEC) repository documents typical attack patterns employed by adversaries to exploit system weaknesses.
%The link between vulnerability and attack pattern is explicit only in rare cases.
%In contrast, correlating these data sources to identify relationships between specific vulnerabilities and attack patterns is a laborious task that requires cybersecurity operators to interpret and link the data manually.
Direct connections between vulnerabilities and attack patterns are crucial for enhanced threat analysis~\cite{10.1007/978-3-031-70903-6_12}, forensics~\cite{inam2022faust,touloumis2022tool}, and cyber risk assessment and response~\cite{chi2023prioritizing,hankin2022attack}.
Nonetheless, these connections are explicitly documented only in limited cases, and establishing meaningful relationships between these data sources remains a challenging and time-consuming process, often requiring strong manual effort~\cite{teuwen2025ruling,yadav2024r+}.
% cybersecurity professionals to interpret and link information manually~\cite{teuwen2025ruling,yadav2024r+}.

Previous efforts to address this issue have employed information retrieval techniques~\cite{kotenko2015capec,ruohonen2018toward,xianghui2015research} and ontology-based mappings~\cite{lee2020practical,stellios2021assessing,zhu2019research,ghosh2025cve}.
More recent studies leverage Natural Language Processing (NLP) to uncover semantic relationships between textual descriptions of vulnerabilities and attack patterns~\cite{giannakopoulos2023usage,mdpi2022capeccve,10803510}.
While these methods offer promising steps toward automation, they often struggle to incorporate contextual nuances and domain-specific terminology, such as acronyms and critical keywords, that are vital in cybersecurity.

This paper addresses these limitations by introducing \oursystem, an NLP-based methodology that integrates semantic similarity with keyword analysis to improve the accuracy of CVE-CAPEC associations.
%Unlike purely semantic approaches, our proposal incorporates keyword analysis to capture domain-specific terms, critical to identifying precise relationships.
%The proposed methodology balances semantic contextual understanding with specific technical details by incorporating both techniques.
Unlike approaches that rely solely on semantic similarity, \oursystem emphasizes domain-relevant terminology, enabling accurate relationship identification. Combining semantic understanding and targeted keyword search, the proposed methodology combines contextual and technical insights.
To assess its effectiveness, we conduct extensive experiments using literature-derived and manually expanded datasets.
Our method is evaluated against state-of-the-art models and demonstrates superior performance in linking vulnerabilities with attack patterns.
Furthermore, we introduce a novel dataset of CVE-CAPEC associations, three times larger than those used in previous works, curated and validated by four security experts. This dataset is publicly released to support future research.
Ultimately, \oursystem enhances automated vulnerability-threat analysis, reducing dependence on manual efforts and enabling cybersecurity professionals to focus on higher-priority tasks.

\section{Background and Problem Statement}
\label{sec:background}

% \emph{Common Vulnerabilities and Exposures} (CVE)~\cite{cve_link} is a dictionary of publicly disclosed cybersecurity vulnerabilities and exposures managed by the MITRE corporation.
% CVE~\cite{cve_link} is a dictionary of publicly disclosed cybersecurity vulnerabilities and exposures managed by the MITRE corporation, where entries are collected, organized, and published in other repositories such as the \emph{National Vulnerability Database} (NVD)~\cite{nvd_link}, managed by NIST.
CVE\footnote{\url{https://www.cve.org/}} is a MITRE-maintained dictionary of publicly disclosed vulnerabilities, where entries are collected, organized, and published in other repositories such as the \emph{National Vulnerability Database} (NVD)\footnote{\url{https://nvd.nist.gov/}}.
% The CVE dictionary consists of a set of independent entries, each of which has (i) an identification number, (ii) a description, and (iii) at least one public reference to external sources providing information about the issue.
% A CVE identifier follows a simple syntax: \texttt{CVE}-\texttt{\emph{YEAR}}-\texttt{\emph{5digits}}, where \texttt{\emph{YEAR}} indicates when the identifier was created and \texttt{\emph{5digits}} is a sequence number in the year. When someone suspects a vulnerability in a piece of software, s/he requests an identifier from MITRE. MITRE assigns the identifier (\texttt{\emph{YEAR}} and \texttt{\emph{5digits}}) but does not publish it immediately. The identifier is only published when the vulnerability is confirmed. In some cases, this process may take a significant amount of time, and some identifiers are never confirmed and thus never published.
% 
% In addition to the dictionary maintained by MITRE, CVE entries are collected, organized, and published in other repositories such as the \emph{National Vulnerability Database} (NVD)~\cite{nvd_link}, managed by NIST.
% NVD is built upon and fully synchronized with the CVE list so that any update to CVE appears in NVD.
%
Each CVE entry has an identifier, description (i.e., a textual description of the vulnerability), relevant references (e.g., to software weaknesses from the \emph{Common Weakness Enumeration} (CWE) repository\footnote{\url{https://cwe.mitre.org/}}, affected software configurations, and metrics associated with CVSS\footnote{\url{https://www.first.org/cvss/}}).
% 
% is structured as follows: 
% \begin{itemize}
% \item \emph{CVE identifier}: it is the unique identifier associated with the vulnerability.
% \item \emph{CVE assigner}: it indicates the entity that reported the vulnerability.
% \item \emph{CVE description}: it is a textual description of the vulnerability and the effect of a possible exploit.
% \item \emph{Relevant References}: it provides a set of references that enrich the knowledge about the vulnerability context, including:
% 	\begin{enumerate*}[label=(\roman*)]
% 		\item Reference to software weaknesses classified in the \emph{Common Weakness Enumeration} (CWE) repository~\cite{cwe_link}.
% 		\item Reference to known affected software configurations that assigns to the CVE one (or more) \emph{Common Platform Enumeration} (CPE) string(s) used to identify information technology systems, software, and packages affected by the vulnerability.
% 		\item A severity score and the associated CVSS vector. 
%         % \simo{fa parte delle relevant references? @Ale: si}
% 	\end{enumerate*}
% \item \emph{CVE history}: temporal information related to vulnerability history, such as the date of publication and the date of the last update.
% \end{itemize}

%\subsection{Common Attack Pattern Enumeration and Classification (CAPEC)}
% Meanwhile, another relevant knowledge base is \emph{Common Attack Pattern Enumeration and Classification} (CAPEC)~\cite{capec_link},
CAPEC\footnote{\url{https://capec.mitre.org/}} is a publicly available catalog of common attack patterns reporting on how adversaries may exploit weaknesses or flaws in applications. 
% It is managed by MITRE and maintained by a community project.
% 
% The CAPEC data comprises \emph{views}, \emph{categories}, and \emph{attack patterns}. 
An \emph{attack pattern} is a generalized description of how adversaries execute a specific phase of an attack. Each entry includes common characteristics, challenges faced by attackers, and mitigation strategies. 
They are derived from detailed analyses of real-world exploits.
% and are intended to inform both detection and defense mechanisms.
% The CAPEC repository is organized in a graph-based structure with concepts classified in \emph{views}, \emph{categories}, and \emph{attack patterns}. 
% An \emph{attack pattern} entry specifies common attributes characterizing an attack and approaches employed by adversaries to exploit known weaknesses. 
% Attack patterns also define the challenges that an adversary may face and are generated from in-depth analysis of specific real-world exploit examples.
% Each attack pattern captures knowledge about how specific parts of an attack are designed and executed and guides how to mitigate the attack's effectiveness.
CAPEC organizes attack patterns into hierarchical relationships, where complex or broad attack patterns may serve as \emph{parent} to more specific \emph{child} patterns. This structure allows for a layered understanding of attacks, enabling analysts to trace high-level strategies down to concrete implementations.
% % 
% %\emph{View} and \emph{category} entries are built with the following rationale: \emph{view} nodes identify a perspective to support the traversal of the CAPEC repository, while \emph{category} nodes model high-level concepts that serve as a point of aggregation for attack patterns.
% \emph{View} and \emph{category} provide additional conceptual scaffolding. \emph{View} offers different perspectives for traversing the repository, while \emph{category} aggregates related attack patterns under broader conceptual groupings. This organization facilitates both high-level analysis and fine-grained exploration of adversarial techniques.

\emph{Common Weakness Enumeration} (CWE) repository is a taxonomy of software and hardware weakness types used to standardize the classification of vulnerabilities and support detection and prevention activities.

%\subsection{Problem Statement}
\noindent{\bf Problem Statement.}
% %\sil{qui va specificato chiaramente che non si puo' fare il collegamento CVE - CWE - CAPEC diretto e che questo ha ripercussioni sul trovare dataset grandi}
% Given the current structure of the CVE, CWE, and CAPEC repositories, it is \emph{theoretically} possible to establish conceptual relationships between a specific vulnerability and one or more attack patterns. 
% %To do so, we reference CAPEC entries by CWE entries (i.e., starting from a specific CWE entry, it is possible to understand the related attack patterns), and CWE entries by CVE vulnerabilities (a CWE weakness abstracts and categorizes a generic issue that may be common for different CVE vulnerabilities). Thus, traversing available external references in CVE and CWE entries, it is possible to infer \emph{indirect} relationships between vulnerabilities and attack patterns.  
% This relationship is mediated through the hierarchical linkage of these repositories: CAPEC entries refer to attack patterns and are connected to CWE entries, while CWEs represent classes of weaknesses commonly associated with multiple CVEs. Therefore, by traversing external references from a CVE to its associated CWE(s), and subsequently to CAPEC entries, one can infer \emph{indirect} connections between CVE vulnerabilities and CAPEC attack patterns.
% %However, because of the huge number of CVEs detected and published daily and the analysis effort required to correctly associate CVEs with CWEs, we observe that many vulnerabilities have no associated CWEs and consequently are not linked to any attack pattern for a long time.
Given the structure of the CVE, CWE, and CAPEC repositories, it is possible to infer {\bf indirect links} between vulnerabilities and attack patterns: CAPEC entries describe attack patterns linked to multiple CWEs, which in turn represent weakness classes tied to multiple CVEs.
The main issue with this structure is that it is often incomplete in practice. Establishing and maintaining CVE–CWE mappings is a labor-intensive process~\cite{angelini2020toward}. As a result, a significant number of vulnerabilities lack associated CWE classifications, which in turn prevents any inferred linkage to relevant attack patterns in CAPEC, as proved empirically by the ``NVD Data Feeds''. 
It reports that the percentage of CVEs with at least one referenced CWE was extremely low in the early years, with a peak of only around $20\%$ in 2003\footnote{\url{https://nvd.nist.gov/vuln/data-feeds}}.

Moreover, even when CWE references are present, the indirect nature of the CVE–CWE–CAPEC connection often results in generalized associations that may be too abstract for practical, real-world threat assessments. CWEs typically represent high-level weaknesses, which do not always map precisely to the specific techniques or intentions behind a given attack pattern.
For example, CVE-2021-22986 (a vulnerability in F5 BIG-IP/iControl REST) is associated with CWE-20 (Improper Input Validation), a very broad category. This CWE corresponds to many CAPEC entries, such as CAPEC-137 (Parameter Injection), CAPEC-138 (Command Injection), or CAPEC-111 (HTTP Response Splitting). Without additional context or semantic analysis, it is unclear which attack pattern applies. 
Consequently, identifying which attack patterns are truly relevant for a given vulnerability remains non-trivial. 
% This hampers the timely and accurate assessment of potential threats.
%This paper tackles the problem of supporting threat analysis with a time-effective mechanism to detect relevant connections between CVE vulnerabilities and CAPEC attack patterns. 
%More in detail, given a set of CVE vulnerabilities $V$ and the set of all possible attack patterns $P$ extracted from CAPEC, we want to automatically estimate how much a vulnerability $v \in V$ is connected to/potentially enables an attack pattern $p \in P$.
%This enables improved and more comprehensive analysis of threats and vulnerabilities as CAPEC provides additional information otherwise not included in the CVE knowledge base.

This paper proposes an efficient method for automatically detecting links between CVE vulnerabilities and CAPEC attack patterns, enriching CVE data with actionable threat insights to improve the depth and accuracy of threat modeling.
\section{Related Work}
\label{sec:related}

Different works in the literature investigate the integration of various cybersecurity knowledge bases for comprehensive threat analysis~\cite{kaloroumakis2021toward,nair2022mapping}.
Among them, some solutions propose the usage of ontologies like Zhu et al.~\cite{zhu2019research}, who standardize the relationships between vulnerabilities and weaknesses, and Stellios et al.~\cite{stellios2021assessing}, who develop a methodology to identify critical attack paths in cyber-physical systems.
Similarly, Lee et al.~\cite{lee2020practical} introduce a formal system for cyber kill chain-based analysis, while Ghosh et al.~\cite{ghosh2025cve} use NLP to learn vulnerability evaluation from the vulnerability history of medical devices to build an ontology portfolio.
Other approaches leverage information retrieval~\cite{kotenko2015capec,ruohonen2018toward}, such as Xianghui et al.~\cite{xianghui2015research}, who leverage the relationships between CAPEC, CWE, and CVE databases for improved threat analysis.

While these approaches help analyze cyber threats from different perspectives (e.g., risks and countermeasures), analysts still require a great effort to correlate distinct cybersecurity knowledge bases.
For this reason, a recent trend leverages NLP techniques to explore the relationships between cybersecurity knowledge bases.
% 
% For example, different works investigate models to correlate vulnerabilities from CVE to concepts from MITRE ATT\&CK~\cite{ampel2021linking,kuppa2021linking}.
Some works aim to identify possible vulnerability-associated tactics for a more informed threat analysis~\cite{kuppa2021linking}.
Sun et al.~\cite{sun2021generating} introduce a method for extracting details from ExploitDB --a publicly available repository of exploits-- and generating CVE descriptions through supervised ML, thus requiring a significant effort to annotate the necessary entries.
With a different model, Silvestri et al.~\cite{silvestri2023machine} propose an NLP methodology based on BERT~\cite{bert2019devlin} to extract vulnerability details in the healthcare sector. In contrast, Mounika et al.~\cite{mounika2019analyzing} use unsupervised topic modeling to analyze CVE vulnerabilities.
All these studies focus on facilitating the analysis of vulnerability descriptions from the CVE database but do not extensively explore their connections to other cybersecurity knowledge bases, particularly the CAPEC attack patterns.
 
\review{Two works are especially relevant in this context.
% Othman et al.~\cite{10803510} examine five feature extraction methods (TF-IDF, LSI, BERT, MiniLM, and RoBERTa) for a preliminary investigation of their capability to associate vulnerabilities with their attack patterns.
Kanakogi et al.~\cite{mdpi2022capeccve} propose the usage of NLP to identify CAPEC attack patterns based on CVE descriptions, compare the similarity of CAPEC descriptions with CVE ones, and rank them based on it to find the most suitable CAPEC.
They provide models more performant than existing preliminary investigations~\cite{10803510}.
Differently, Giannakopoulos and Maliatsos~\cite{giannakopoulos2023usage} proposed a supervised approach that trains an ML model with a privately labeled dataset to associate CVE vulnerabilities with general threats.
Both approaches focus on investigating and using state-of-the-art models to associate attack patterns with vulnerabilities.
In contrast, we propose a novel approach that combines semantic similarity with keyword search to improve the capabilities of these models to classify associations accurately.
We compare our solution with these approaches~\cite{giannakopoulos2023usage,mdpi2022capeccve} and show that the proposed one performs better.}{
Two works are particularly related to our problem setting, but differ in scope and methodology. 
Kanakogi et al.~\cite{mdpi2022capeccve} evaluate generic NLP similarity pipelines for tracing CAPEC attack patterns from CVE descriptions, comparing document representations and off‑the‑shelf embedding models with cosine similarity.
In parallel, Giannakopoulos and Maliatsos~\cite{giannakopoulos2023usage} train supervised models on a privately labeled dataset to associate CVE entries with higher‑level threat categories, focusing on classification performance in a threat‑level taxonomy rather than on fine‑grained CAPEC mappings.
In contrast, this paper adopts a different perspective. 
Instead of relying solely on generic similarity pipelines or purely supervised threat classification, \oursystem is a new methodology explicitly designed as a hybrid, domain‑specific scoring framework for CVE--CAPEC association, moving beyond best‑pipeline selection to a scoring architecture in which domain terminology systematically modulates the contribution of semantic similarity.}
\section{Methodology design and implementation}
\label{sec:methodology}

\begin{figure}[h!]
    \centering
    \includegraphics[width=\linewidth]{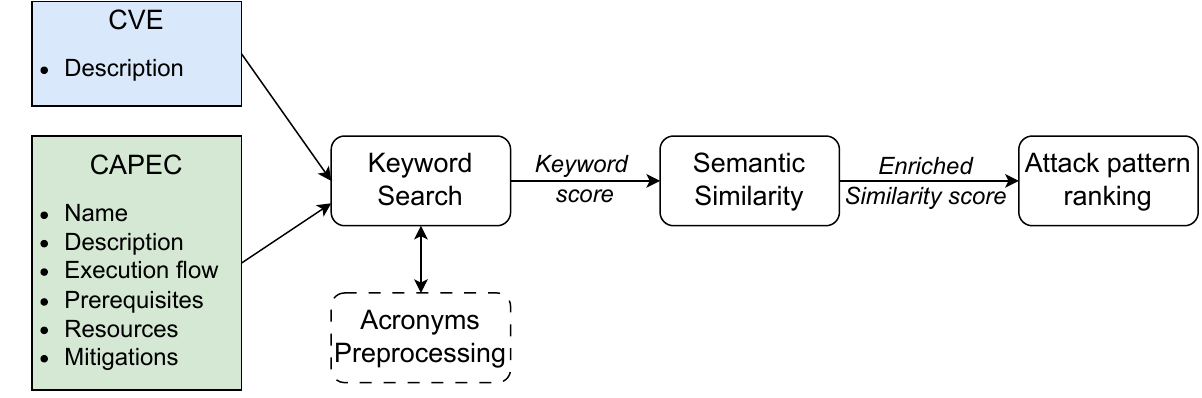}
    \caption{\oursystem methodology overview.}
    \label{fig:approach_overview}
\end{figure}

\review{In this section, we present \oursystem to estimate the degree of relevance of a single (or multiple) CVE vulnerability for a set of attack patterns by using NLP techniques.
The idea considers estimating the degree of relevance of a CVE for an attack pattern by comparing the description of the vulnerability with information describing the attack pattern to verify whether there is a match between concepts. 
To this aim, we measure the degree of relevance as a function of the similarity between the vulnerability and what is needed to perform the attack pattern.}{%
In this section, we present \oursystem, which estimates the degree of relevance of CVE vulnerabilities to a set of attack patterns by leveraging NLP.
The core idea is to assess how relevant a CVE is to a given attack pattern by comparing the vulnerability description with the textual information that characterizes the attack pattern to identify overlapping or closely related concepts.
The degree of relevance is modeled as a function of the similarity between the vulnerability description and the information required to perform the attack pattern.}

\review{As an example, vulnerability CVE-2002-0412 description states 
``[...] allows remote attackers to execute arbitrary code by causing format strings to be injected into calls to the syslog function''.
Looking at CAPEC, we can identify the attack pattern CAPEC-67 in which the first step of the execution flow states ``[...] In this attack, adversaries look for applications that use syslog() incorrectly''.
The high similarity of these attributes suggests a strong association between CVE-2002-0412 and CAPEC-67, which can be used for more informative threat analysis, such as associating the different attack steps from CAPEC to the vulnerability from CVE (this information would not otherwise be retrievable from the only CVE repository).
\Cref{fig:approach_overview} reports the overview of \oursystem.}{%
As an illustrative example, the description of vulnerability CVE-2002-0412 states \textit{``[...] allows remote attackers to execute arbitrary code by causing format strings to be injected into calls to the syslog function''}. Within CAPEC, the attack pattern CAPEC-67 includes, as part of its execution flow, the statement \textit{``[...] In this attack, adversaries look for applications that use syslog() incorrectly''}.
The strong conceptual overlap between these descriptions suggests a close association between CVE-2002-0412 and CAPEC-67, useful for more informative threat analysis.
% , for instance by linking the attack execution steps described in CAPEC to the vulnerability documented in CVE; information that would not be directly retrievable from the CVE repository alone.
}

\review{Our estimation starts by collecting vulnerability and attack pattern information from CVE and CAPEC repositories.
We identify two main analytical tasks (see \Cref{fig:approach_overview}): keyword search and semantic similarity.
\textbf{Keyword search} models the exact or near-exact matches of words within a text~\cite{firoozeh2020keyword}, thus looking for specific terms or combinations of terms in the text.
The rationale for keyword search is that vulnerability descriptions from CVE often contain specific terminology that aligns directly with CAPEC patterns. 
By identifying these keywords, the correlation benefits from a clear and explicit link between the vulnerability and the relevant attack pattern, which helps ground the model's predictions more reliably.
The knowledge from the keyword search enriches the information commonly derived by semantic similarity.
\textbf{Semantic similarity} measures how similar the textual information of vulnerabilities (from CVE) and attack patterns (from CAPEC) are in meaning, regardless of their exact words or phrasing~\cite{chandrasekaran2021evolution}.
While semantic similarity models lend considerable help in the task of associating vulnerabilities to attack patterns, keyword search further improves their performance by considering context-specific concepts such as attack names or specific threat exploits.
In fact, semantic similarity and keyword search are combined to allow a comprehensive correlation between CVE and CAPEC that finally results in the computation of \textbf{attack pattern ranking} to identify the most appropriate attack patterns for each vulnerability.}{%
\Cref{fig:approach_overview} provides an overview of \oursystem.
Our methodology starts by collecting vulnerability and attack pattern information from the CVE and CAPEC repositories.
We identify two main analytical tasks (see \Cref{fig:approach_overview}): keyword search and semantic similarity.
\textbf{Keyword search} captures exact or near-exact matches of terms within textual data~\cite{firoozeh2020keyword}, focusing on the explicit presence of specific words or combinations of words.
The motivation for keyword search is that CVE descriptions often include highly specific technical terminology that directly aligns with CAPEC attack pattern names or labels.
Identifying these terms provides a clear and explicit signal of correlation, grounding the association between vulnerabilities and attack patterns.
The information derived from keyword search complements and enriches the evidence provided by semantic similarity.
\textbf{Semantic similarity} measures how similar the vulnerability description from CVE and the attack pattern information from CAPEC are in meaning, independently of their exact wording or phrasing~\cite{chandrasekaran2021evolution}.
While semantic similarity models are effective in capturing contextual and conceptual relations, keyword search further improves performance by accounting for domain-specific expressions such as well-known attack names or exploit labels.
Together, semantic similarity and keyword search are combined to produce an \textbf{attack pattern ranking}, which identifies the most relevant attack patterns for each vulnerability.}

\subsection{Input data}
\label{ssec:input}
\review{To associate vulnerabilities with attack patterns, the inputs of \oursystem are the list of CVE vulnerabilities that must be analyzed (e.g., the ones identified by vulnerability scanners of a certain network) and the collection of all the non-deprecated CAPEC attack patterns (currently $559$).}{%
To associate vulnerabilities with attack patterns, the inputs of \oursystem consist of a set of CVE vulnerabilities to be analyzed (e.g., those identified by vulnerability scanners within a given network) and the collection of all non-deprecated CAPEC attack patterns (currently $559$).}

\review{As mentioned by Kanagozi et al.~\cite{mdpi2022capeccve}, relevant vulnerability information from CVE is included almost exclusively in the textual \emph{description} attribute, thus representing the main information source. 
In contrast, CAPEC presents more explanatory attributes: we leverage \emph{name} and \emph{description} as they include high-level details and the main source of textual information.
We include also \emph{attack execution flow} and \emph{mitigations}. The former is a step-by-step explanation of how the attack is carried out, typically presented in the form of prerequisites and attack steps; the latter is a description of the recommended measures or strategies for the attack pattern.
Finally, we consider the \emph{prerequisites} and \emph{resources} attributes that describe the detailed conditions for the attack to be executed and the tools, technologies, or resources needed for the attack.}{%
As observed by Kanagozi et al.~\cite{mdpi2022capeccve}, relevant vulnerability information in CVE is conveyed almost exclusively through the textual \emph{description} attribute, which therefore represents the primary source of information.
In contrast, CAPEC provides a richer set of explanatory attributes.
In our approach, we leverage the \emph{name} and \emph{description} fields, as they capture high-level characteristics of the attack pattern and constitute the main textual sources.
We also include the \emph{attack execution flow} and \emph{mitigations} attributes.
The former describes the attack as a sequence of prerequisites and execution steps, while the latter outlines recommended defensive measures.
Finally, we consider the \emph{prerequisites} and \emph{resources} attributes, which detail the conditions under which the attack can be carried out and the tools or technologies required.}

\subsection{Keyword search}
\label{ssec:keyword}

\review{The aim of keyword search is to verify if the same keywords are in both the CVE and CAPEC attributes.
If this happens, we have an additional indication to strengthen the estimation and thus refine the similarity score.
The keyword search module evaluates the connection between CVE vulnerability descriptions and CAPEC attack patterns by identifying shared keywords. 
It takes as input the CVE description, the CAPEC name, and a list of acronyms, while the output is a keyword-based relevance score.}{%
The goal of keyword search is to determine whether relevant keywords appear in both CVE and CAPEC textual attributes.
When such overlaps are found, they provide additional evidence that strengthens the estimated association and refines the overall relevance score.
The keyword search module evaluates the relationship between CVE vulnerability descriptions and CAPEC attack patterns by identifying shared keywords.
Its inputs are the CVE description, the CAPEC name, and a predefined list of acronyms, while its output is a keyword-based relevance score.}

\review{To achieve this goal, we perform an \emph{acronyms preprocessing} step.
It consists of the definition of a mapping of terms with their acronyms.
For example, in CVE and CAPEC knowledge bases, ``SQLi'' replaces ``SQL Injection'' very often; similarly, ``XSS'' for Cross-Site Scripting.
These acronyms are closely tied to specific attack patterns and might be missed by a purely semantic approach.
For example, \textit{CVE-2016-1000113} is described as \textit{``XSS and SQLi in huge IT gallery v1.1.5 for Joomla''}.
In this case, the acronyms XSS (Cross-Site Scripting) and SQLi (SQL Injection) provide direct indicators of the types of vulnerabilities present. 
While a semantic similarity model might capture the general notion of web-based vulnerabilities, the presence of these acronyms offers precise clues to identify specific CAPEC attack patterns, such as \textit{CAPEC-66 (SQL Injection)} or \textit{CAPEC-63 (Cross-Site Scripting)}.}{%
To support this process, we first perform an \emph{acronyms preprocessing} step.
This step defines a mapping between technical terms and their commonly used acronyms.
For instance, in both CVE and CAPEC repositories, ``SQLi'' is frequently used in place of ``SQL Injection'', and ``XSS'' is used for ``Cross-Site Scripting''.
These acronyms are strongly associated with specific attack patterns and may not be fully captured by semantic similarity alone.
For example, \textit{CVE-2016-1000113} is described as \textit{``XSS and SQLi in huge IT gallery v1.1.5 for Joomla''}.
Here, the acronyms XSS and SQLi directly indicate the vulnerability types involved.
While semantic similarity may identify a general class of web-related vulnerabilities, the explicit presence of acronyms enables a precise association with specific CAPEC attack patterns, such as \textit{CAPEC-66 (SQL Injection)} or \textit{CAPEC-63 (Cross-Site Scripting)}.}

\review{The mapping between acronyms and their expansions has been collected from the NIST glossary\footnote{\url{https://csrc.nist.gov/glossary}} and augmented with state-of-the-art on systems and network security~\cite{scarfone2010system}.
Other technical terms present in security knowledge bases and not reported by existing documentation have been manually added\footnote{The collection of acronyms used in the proposed approach is available in the open-source repository.}.}{%
The acronym-to-term mapping is derived from the NIST glossary~\footnote{\url{https://csrc.nist.gov/glossary}} and extended using state-of-the-art security literature~\cite{scarfone2010system}.
Additional technical terms commonly used in security knowledge bases but not covered by existing references were manually included~\footnote{The complete list of acronyms is available in the open-source repository.}.}

\review{The keyword search algorithm first extracts the keywords from the CAPEC \textit{name} and \textit{alternate\_terms} attributes.
This is done with an algorithm inspired by the existing keyword extraction approaches~\cite{firoozeh2020keyword,rose2010automatic} and working as follows:}{%
The keyword search algorithm begins by extracting keywords from the CAPEC \textit{name} and \textit{alternate\_terms} attributes, following a procedure inspired by established keyword extraction methods~\cite{firoozeh2020keyword,rose2010automatic}:}

\noindent \textbf{Step 1: strings preprocessing.} 
\review{Text strings from CVE and CAPEC are preprocessed by removing spaces, tabs, punctuation, uppercase characters, and stop words (i.e., common words such as conjunctions, prepositions, and articles) that do not carry significant meaning in the context.}{%
Textual descriptions from CVE and CAPEC are preprocessed using standard techniques like lowercasing, removal of punctuation and extraneous whitespace, tokenization, and elimination of stop words (i.e., high-frequency function words such as articles, conjunctions, and prepositions contributing little semantic information).}

\noindent\textbf{Step 2: lemmatization.}
\review{The text is then lemmatized~\cite{khyani2021interpretation}, reducing words to their base or dictionary form to ensure consistency. 
This step is crucial for aligning variations of the same word and simplifying the task of keyword matching.}{%
The resulting tokens are lemmatized~\cite{khyani2021interpretation}, reducing words to their canonical dictionary form.
This ensures that different inflected forms of the same word are treated consistently during matching.}

\noindent\textbf{Step 3: acronyms preprocessing.}
\review{We look for acronyms in vulnerability and attack pattern descriptions.
In this way, if the attack pattern name fully represents the acronym (e.g., \textit{CAPEC-63: Cross-Site Scripting (XSS)}), it receives the maximum keyword score because it fully represents an entire concept or attack.
On the contrary, if it contains the acronym but is not a full expansion, the ``extra'' words are searched within the vulnerability description.
For example, \textit{CAPEC-592: Stored XSS} does not refer precisely to XSS but to its side storage concept. In that case, the term ``stored'' is searched, and if present in the vulnerability description, the association receives the maximum values.
The rationale of this empirical choice is to balance similarity and keyword scores for the final score.
In fact, such a value is the maximum one that is granted to represent scores in the range $[0, 1]$.
On the contrary, if the attack pattern name does not have corresponding acronyms in the vulnerability descriptions, we follow common text processing approaches~\cite{firoozeh2020keyword} to check the number of keywords commonly present in the attack pattern and vulnerability attributes.}{%
The algorithm searches for acronyms within both vulnerability descriptions and attack pattern names.
If the CAPEC name fully expands an acronym present in the CVE description (e.g., \textit{CAPEC-63: Cross-Site Scripting (XSS)}), the association is assigned the maximum keyword score, as the acronym represents a complete attack concept.
If the CAPEC name contains the acronym in a more specific form (e.g., \textit{CAPEC-592: Stored XSS}), the additional qualifying terms (such as ``stored'') are searched for in the CVE description.
If they are present, the maximum keyword score is again assigned.
When no relevant acronyms are identified, the algorithm follows conventional text processing approaches~\cite{firoozeh2020keyword} to identify the number of shared keywords between the CAPEC attributes and the CVE description.}

\noindent\textbf{Step 4: alternate terms processing.} 
\review{Some attack patterns present the \textit{``alternate\_terms''}, which contain alternative terminologies for describing the same attack pattern. 
In these cases, we repeat the keyword search for each term, and the keyword score is the maximum value between the scores obtained for the name field and each value of the alternate terms.}{%
When the CAPEC attack pattern includes \textit{``alternate\_terms''}, they are processed independently using the same keyword extraction procedure.
The keyword score for the attack pattern is then defined as the maximum score among the name and all alternate terms.}

\noindent\textbf{Step 5: score computation.} 
\review{When the keywords are processed, the last step of the algorithm computes the keyword scores.
The final output of the keyword search module is a \textit{keyword score}. The keyword score is the ratio of the identified keywords to the total number of keywords, and therefore, it is in the range $[0, 1]$.
The rationale is to quantify how much the CAPEC attack pattern is reflected in the CVE description by measuring the presence and relevance of shared keywords and acronyms. 
This score would indicate the strength of alignment between the CVE description and the CAPEC title, suggesting whether the CAPEC attack pattern might be explicitly mentioned or referenced in the CVE vulnerability.}{%
In the final step, the keyword score is computed as the ratio between the number of identified shared keywords (including matched acronyms) and the total number of extracted keywords.
As a result, the keyword score lies in the range $[0,1]$.
This score quantifies the extent to which the CAPEC attack pattern is explicitly reflected in the CVE description and provides a direct measure of their terminological alignment.}

\subsection{Semantic similarity}
\label{ssec:similarity}

\review{The semantic similarity module quantifies the similarity between CVE vulnerability descriptions and CAPEC attack pattern attributes.
Semantic similarity refers to measuring how similar two pieces of text are in meaning, regardless of their exact words or phrasing. It involves understanding the context and the relationships between words, phrases, sentences, or larger pieces of text~\cite{chandrasekaran2021evolution}.
It is meaning-focused because it assesses contextual similarity rather than specific keywords and handles synonyms, paraphrasing, and semantic relations.}{%
The semantic similarity module quantifies the conceptual similarity between CVE vulnerability descriptions and CAPEC attack pattern attributes.
Semantic similarity focuses on the meaning of text rather than exact word matches, capturing contextual relationships, paraphrasing, and synonyms~\cite{chandrasekaran2021evolution}.}

\review{We select two main models that currently show the best performance to perform semantic similarity for cybersecurity-related tasks: SBERT and ATT\&CK BERT. 
\emph{SBERT} (SentenceTransformer)~\cite{reimers2019sentence} is a variant of BERT~\cite{bert2019devlin} for efficient computation of sentence similarity. 
A text string is the input to the model, which divides it into smaller units called \emph{tokens}. These tokens represent individual (parts of) words and are transformed into \emph{embedding vectors}, capturing their meaning and relationship to other tokens. 
The embeddings are combined into a single vector representing the entire sentence. In this way, the semantic similarity between sentences is traced back to the geometric distance between vectors: the smaller the distance, the greater the similarity.
The second model we employ in the proposed solution is \emph{ATT\&CK BERT}~\cite{10.1007/978-3-031-37586-6_15}\footnote{\url{https://huggingface.co/basel/ATTACK-BERT}}. 
It has a similar function to SBERT, finetuned for the cybersecurity domain.
In fact, the NLP model has been trained on specific cybersecurity datasets (e.g., 
MITRE ATT\&CK~\cite{abdeen2023smet}), including texts related to vulnerabilities, attacks, and exploit techniques.}{%
We adopt two state-of-the-art models that have demonstrated strong performance in cybersecurity-related NLP tasks: SBERT and ATT\&CK BERT.
\emph{SBERT} (SentenceTransformer)~\cite{reimers2019sentence} is a BERT-based model optimized for computing sentence-level embeddings.
Given a text string, the model tokenizes the input and maps it to a fixed-length embedding vector that represents its semantic content.
Semantic similarity between two texts is then computed based on the geometric relationship between their embedding vectors: the closer the vectors, the higher the similarity.
The second model is \emph{ATT\&CK BERT}, a variant of BERT fine-tuned on cybersecurity-specific corpora, including MITRE ATT\&CK~\cite{abdeen2023smet}.
This domain adaptation allows the model to better capture security-related terminology and concepts.}

\review{The state-of-the-art metric to measure the similarity between word embeddings is \emph{cosine similarity}, as it represents well the distance between them in terms of the cosine of the angle between two vectors~\cite{lahitani2016cosine}.
Cosine similarity is defined as $\cos(\theta) = \frac{\mathbf{X} \cdot \mathbf{Y}}{\|\mathbf{X}\| \|\mathbf{Y}\|}$, where $\mathbf{X}$ and $\mathbf{Y}$ are the sentence embeddings generated by the model, and \(\mathbf{X} \cdot \mathbf{Y}\) is their dot product.
\(\|\mathbf{X}\|\) and \(\|\mathbf{Y}\|\) are the norms of vector \(\mathbf{X}\) and \(\mathbf{Y}\)~\cite{lahitani2016cosine}.
Cosine similarity is within the $[-1, 1]$ range, where the upper extreme indicates that the vectors have the same direction (maximum similarity). In contrast, the lower extreme indicates that the vectors have opposite directions (minimum similarity).
According to cosine similarity, we compute \emph{similarity scores} between the description of the vulnerability from CVE and each selected field of the attack pattern from CAPEC.
In this way, we have $n$ similarity scores $s_1(v,ap), \ldots, s_n(v,ap)$ for each pair of vulnerability $v$ and attack pattern $ap$.
We evaluate the similarity score of a vulnerability $v$ with respect to an attack pattern $ap$ as the mean of its scores.
More formally:
$
    s(v,ap) = \frac{\sum_{i=1}^{n} s_i(v,ap)}{n},
$
    \label{eq:similarity}
where $n$ is the number of attack pattern attributes.
It represents an aggregate normalized measure of the semantic similarity between the vulnerability description and the attack pattern, where each attribute contributes evenly to the overall score by using the mean.
It returns a $[0, 1]$ value.}{%
For both models, similarity between embeddings is computed using \emph{cosine similarity}, a widely adopted metric for measuring the angular distance between vectors~\cite{lahitani2016cosine}.
Cosine similarity is defined as
$
\cos(\theta) = \frac{\mathbf{X} \cdot \mathbf{Y}}{\|\mathbf{X}\| \|\mathbf{Y}\|},
$
where $\mathbf{X}$ and $\mathbf{Y}$ are the embedding vectors of the textual contents.
The cosine similarity value lies in the interval $[-1,1]$, where higher values indicate greater semantic similarity.
In our approach, the CVE description and the selected CAPEC attributes are first vectorized using the chosen model.
Cosine similarity is then computed between the resulting vectors and linearly rescaled to the interval $[0,1]$ to ensure consistency with the keyword score.
This score provides an aggregate, normalized measure of the semantic relatedness between the vulnerability description and the attack pattern.}
% The intermediate similarity scores represent the similarity between the CVE description and a specific field of the CAPEC, for example, the mitigations field. 
% The final similarity score, calculated as the arithmetic mean of the similarity scores of all fields, represents an aggregate, normalized measure of the overall semantic relationship between the CVE description and the different aspects described by CAPEC. Each field contributes evenly to the overall score, providing a balanced picture of the correlation between CVE and CAPEC.
% Eventually, iterating each CAPEC in the process previously described, we then obtain a similarity score for each CVE - CAPEC pair; sorting the scores in descending order we obtain a ranking of CAPECs, from most likely related to least likely related, to the CVE in question.
% The reason for using the arithmetic mean is due to several reasons including the fact that it is an immediate and easy metric to interpret, each field has the same weight in the score (assigning different weights should have a not-so-clear reason), and it also always returns a value between $[0, 1]$.

\subsection{Attack pattern ranking}
\label{ssec:ap_ranking}
\review{The final step of the methodology consists of ranking the attack patterns associated with each vulnerability according to their similarity and keyword scores.
As shown in \Cref{fig:approach_overview}, the proposed solution is a hybrid approach, where we search for CAPEC-related keywords within the CVE description. Then, we compute the similarity scores, and they earn this association a keyword score.
We consider an \textit{overall score} of the CVE association to CAPEC based on the sum of the similarity score and the keyword score, where:
$
    \mathcal{S}_{overall} = \mathcal{S}_{base} + \mathcal{S}_{keyword}.
$
The rationale of the sum is to refine the similarity scores in the presence of specific technical keywords.
For example, \textit{CVE-2006-5288} has description \textit{``[...] have a default administrator username ``root'' and password ``password'' [...]''}.
Its most relevant attack pattern is \textit{CAPEC-70: Try Common or Default Usernames and Passwords}.
The semantic similarity score using SBERT is $0.35$, which is increased to $0.58$ through the keyword search that identifies the keywords ``default'', ``username'', and ``password'' shared by both the CAPEC name and the CVE description.
Based on the overall scores, we define a ranking of the attack patterns from CAPEC for each vulnerability from CVE.
We consider such a rank the driving parameter for the classification, where the position in the ranking is directly proportional to the likelihood of correlation: higher positions indicate attack patterns that are more correlated to the vulnerability. 
The reason why we define rankings is that vulnerability descriptions strongly influence the score values returned by similarity models. 
For example, it happens that some vulnerabilities may have a maximum score for a correlated CAPEC whose value is extremely low compared to other significantly higher ones.
This means that the only values do not appropriately represent the relevance of the associations like the ranking does~\cite{mdpi2022capeccve}.}{%
The final step of the methodology ranks the CAPEC attack patterns associated with each CVE vulnerability based on a combination of keyword search and semantic similarity.
These two components are combined into an \emph{overall score} defined as a weighted sum:
\[
\mathcal{S}_{overall} = \alpha \cdot \mathcal{S}_{semantic} + (1-\alpha) \cdot \mathcal{S}_{keyword},
\]
where $\alpha \in [0,1]$ controls the relative importance of semantic similarity and keyword search.
%In our experiments, we set $\alpha = 0.5$, assigning equal weight to both components.
%The case $\alpha = 0$ corresponds to a configuration in which the keyword search module is not used and the ranking is based solely on semantic similarity.
The rationale for this weighted combination is to refine semantic similarity scores when explicit technical indicators are present.
For example, \textit{CVE-2006-5288} is described as \textit{``[...] have a default administrator username ``root'' and password ``password'' [...]''}.
Its most relevant attack pattern is \textit{CAPEC-70: Try Common or Default Usernames and Passwords}.
While the semantic similarity score alone yields a moderate value, the detection of shared keywords such as ``default'', ``username'', and ``password'' substantially increases the overall score, reinforcing the association.
Based on the overall scores, we generate a ranking of CAPEC attack patterns for each CVE vulnerability.
This ranking serves as the primary output of the system, where higher-ranked attack patterns are considered more strongly correlated with the vulnerability.
We rely on rankings because vulnerability descriptions can vary significantly in length and detail, which may affect raw similarity scores.
% As noted in prior work~\cite{mdpi2022capeccve}, relative ordering provides a more robust and interpretable representation of relevance than absolute scores alone.
}

\section{Evaluation}
\label{sec:evaluation}

\review{We evaluate \oursystem against state-of-the-art methods.}{
We evaluate \oursystem in comparison with state-of-the-art methods, but with a distinct evaluation focus that reflects its different design goals. 
\oursystem is designed as a hybrid, domain‑specific ranking framework for CVE--CAPEC association, combining semantic similarity from SBERT and ATT\&CK‑BERT with an explicit keyword‑based component that captures cybersecurity acronyms, attack names, and other technical terminology as first‑class signals. 
This yields a dedicated scoring architecture in which domain terminology systematically modulates the contribution of semantic similarity, rather than another instance of generic best‑pipeline selection. 
In the following, we assess \oursystem on an expanded, publicly released CVE--CAPEC ground truth and compare its performance with the best configurations reported by Kanakogi et al.~\cite{mdpi2022capeccve} and with the supervised threat‑mapping approach of Giannakopoulos and Maliatsos~\cite{giannakopoulos2023usage}.
}
Code and datasets are available at \url{https://github.com/ds-square/ThreatLinker}.
%\textcolor{red}{Detailed results are provided as supplementary material.}

\subsection{Experimental settings}
\label{sec:exp_settings}

Validating CVE--CAPEC associations requires a reliable Ground Truth (GT), which is currently absent in the literature. 
We thus define two GTs.
The first replicates Kanakogi et al.~\cite{mdpi2022capeccve}, based on CAPEC \textit{Example Instances}, yielding the original $63$ CVE--CAPEC pairs (plus two new associations).
%To provide a larger and more diverse dataset, we created a second GT with $223$ manually validated associations\footnote{\url{https://github.com/Ale96Pa/CVE-CAPEC/blob/main/data/GT2.csv}}.
To provide a larger and more diverse dataset, we created a second GT with $223$ manually validated associations\footnote{\url{https://github.com/ds-square/ThreatLinker/blob/main/cve-capec-mapping.xlsx}}.
\review{Two threat intelligence experts independently classified each CVE--CAPEC pair using descriptions as the main criterion, with two supervisors reviewing all results.}{For each candidate CVE--CAPEC pair, two threat intelligence experts independently assessed whether the association should be considered valid, using the CVE description and multiple CAPEC fields as the primary evidence.
They selected a collection of pseudo-random CVEs so as to diversify the CAPEC patterns covered as much as possible. 
This allowed the construction of a heterogeneous dataset that spans multiple attack categories rather than concentrating on a few dominant ones, making GT2 more suitable for evaluating the robustness of the approach across different types of vulnerabilities and attack techniques.
Disagreements in evaluating associations were resolved through a consensus process involving two supervisors.
The experts first revisited the pair jointly, and unresolved or borderline cases --e.g., associations to very generic CAPEC parents or patterns with only partial semantic alignment-- were escalated to two senior supervisors, who provided a final decision on inclusion or exclusion. 
The construction of this new dataset required detailed analysis of vulnerability descriptions and attack patterns, with particular attention to minimizing annotation errors and avoiding over‑fitting to specific CAPEC families.}
% \review{}{
% 7, aggiungere un paio di supercazzole. 8,  Initially, the selection of CVEs was done completely at random, choosing vulnerabilities distributed between 1999 and 2024, so as to include CVEs whose descriptions could vary in both the language used and the attack techniques. However, most of the random CVEs concerned XSS and SQL Injection, resulting in an unbalanced representation of attacks. For this reason, we changed our strategy by adopting a more structured pseudorandom criterion, i.e. still randomly generating CVEs, over the period 1999-2004, and selecting them so as to diversify the CAPECs covered as much as possible. This made it possible to obtain a more representative and balanced dataset, which was useful for testing the approach on more heterogeneous data. The construction of this new dataset required analysis and careful study of CVE descriptions and various attack patterns, trying to minimise errors in order to obtain a solid ground truth.
% }
% 
\oursystem ranks attack patterns associated with vulnerabilities.
Following standard practice~\cite{tamm2021quality}, we assess performance using \textit{Recall@K}, \textit{Precision@K}, and \textit{Mean Reciprocal Rank} (MRR).
\review{all in the range $[0,1]$.
Recall@K and Precision@K evaluate how accurately relevant attack patterns appear in the top $K$ results, while MRR measures the average position of the first correct association, rewarding systems that rank correct results higher.}{
\textit{Recall@K} is defined as $\frac{\#TP@K}{N}$, where $\#TP@K$ is the number of relevant associations in the first $K$ position, and $N$ is the total number of associations.
\textit{Precision@K} is defined as $\frac{\#TP@K}{K}$.
As a derived metric, \textit{F1-score} is $2 \times \frac{\text{Precision@K} \times \text{Recall@K}}{\text{Precision@K} + \text{Recall@K}}$.
\textit{MRR} is defined as  $\frac{1}{N} \sum_{i=1}^{N} \frac{1}{\text{rank}_i}$, where $N$ is the total number of associations, and $\text{rank}_i$ is the position of the first relevant association in the top-$K$ results.
}
\review{}{We empirically set the weights $\alpha$ to $0.3$ and} compare \oursystem with the current state-of-the-art for CVE--CAPEC association.
% Finally, we study the effects of the CAPEC hierarchical structure on the results obtained by \oursystem.

\subsection{Performance analysis}
We compare \oursystem with Kanakogi et al.~\cite{mdpi2022capeccve}. Since it uses SBERT all-distilroberta-v1 (\emph{SBERT}), SBERT paraphrase-mpnet-base-v2 (\emph{SBERT-mpnet}), and ATTACKBERT, we evaluate \oursystem with the same models for fairness. 
% For fairness, we adopt the same base models in \oursystem.

\subsubsection*{Ground Truth 1}
\begin{figure*}[!th]
    \centering
    \subfloat[Recall@K.\label{fig:recall-k1}]{{\includegraphics[width=0.45\linewidth]{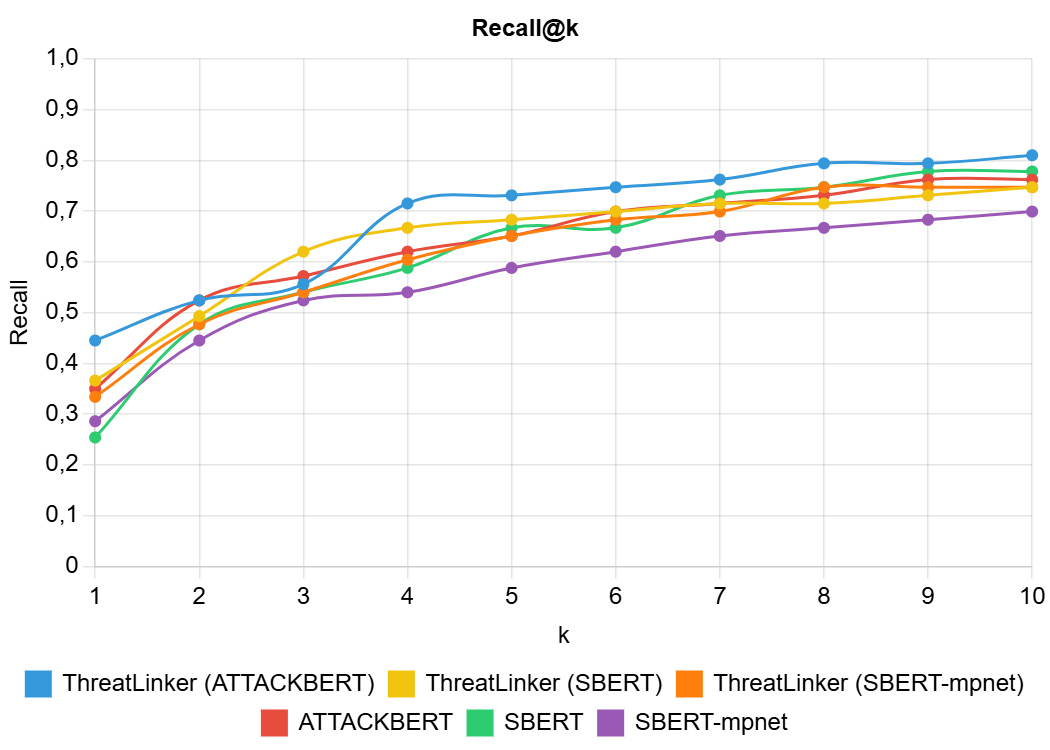}}}
    \qquad
    \subfloat[Precision@K.\label{fig:precision-k1}]{{\includegraphics[width=0.45\linewidth]{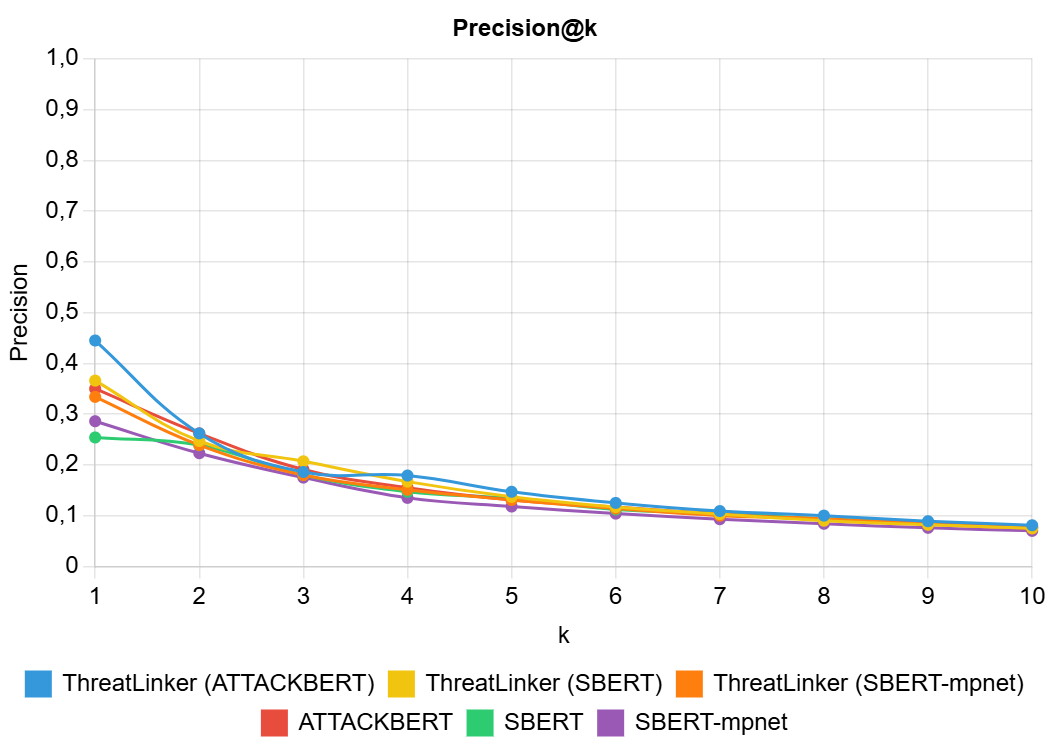}}}
    \caption{Recall@$K$ and Precision@$K$ for GT1~\cite{mdpi2022capeccve}.}
    \label{fig:recall_precision_k1}
\end{figure*}

Using the dataset from~\cite{mdpi2022capeccve}, \Cref{fig:recall_precision_k1} shows Recall@K and Precision@K trends. 
For $K=1$, \oursystem (ATTACKBERT) achieves the best recall ($0.455$), outperforming prior models by $23\%$.
For $K=5$ and $K=10$, recall reaches $0.731$ and $0.81$, respectively, confirming its consistency.
The other models have comparable recall values, as shown by the high overlap.
It is worth noting that the Recall@10, used as the primary metric by~\cite{mdpi2022capeccve}, slightly differs from our experiments ($0.803$ from~\cite{mdpi2022capeccve} against $0.778$ estimated by us).
This change is probably due to the slight differences in text preprocessing and a slightly larger dataset.
Precision@K follows a similar trend, with \oursystem (ATTACKBERT) maintaining the highest precision up to $K=5$.
For values of $K \geq 5$, the Precision@K of the different models are comparable.
%\textcolor{red}{F1-scores (see supplemental material) support these findings.}
% 
\begin{wraptable}{r}{0.25\textwidth}
    %\vspace{-10pt}
    \centering
    \resizebox{0.25\textwidth}{!}{%
    \begin{tabular}{|c|c|}
        \hline
        \textbf{Model} & \textbf{MRR} \\
        \hline  
        \oursystem (ATT.BERT) & \textbf{0.587} \\
        \oursystem (SBERT) & 0.536 \\
        \oursystem (mpnet) & 0.506 \\
        ATTACKBERT & 0.524 \\
        SBERT & 0.462 \\  
        SBERT (mpnet) & 0.457 \\      
        \hline
    \end{tabular}
    }
    \caption{MRR for GT1.}
    \label{tab:mrr-gt1}
    \vspace{-8pt}
\end{wraptable}
MRR values (\Cref{tab:mrr-gt1}) confirm that \oursystem (ATTACKBERT) performs best ($0.587$), followed by \oursystem (SBERT) ($0.536$). 
Overall, \oursystem improves recall, precision, and ranking quality, highlighting its ability to prioritize relevant CVE--CAPEC associations effectively.
\oursystem (ATTACKBERT) is the best-performing model in associating attack patterns with vulnerabilities, according to all the analyzed metrics.
\oursystem improves the performance of all models compared to their classic versions, with a particularly noticeable impact on SBERT.

\subsubsection*{Ground Truth 2}
\begin{figure*}[!ht]
    \centering
    \subfloat[Recall@K.\label{fig:recall-k2}]{{\includegraphics[width=0.45\linewidth]{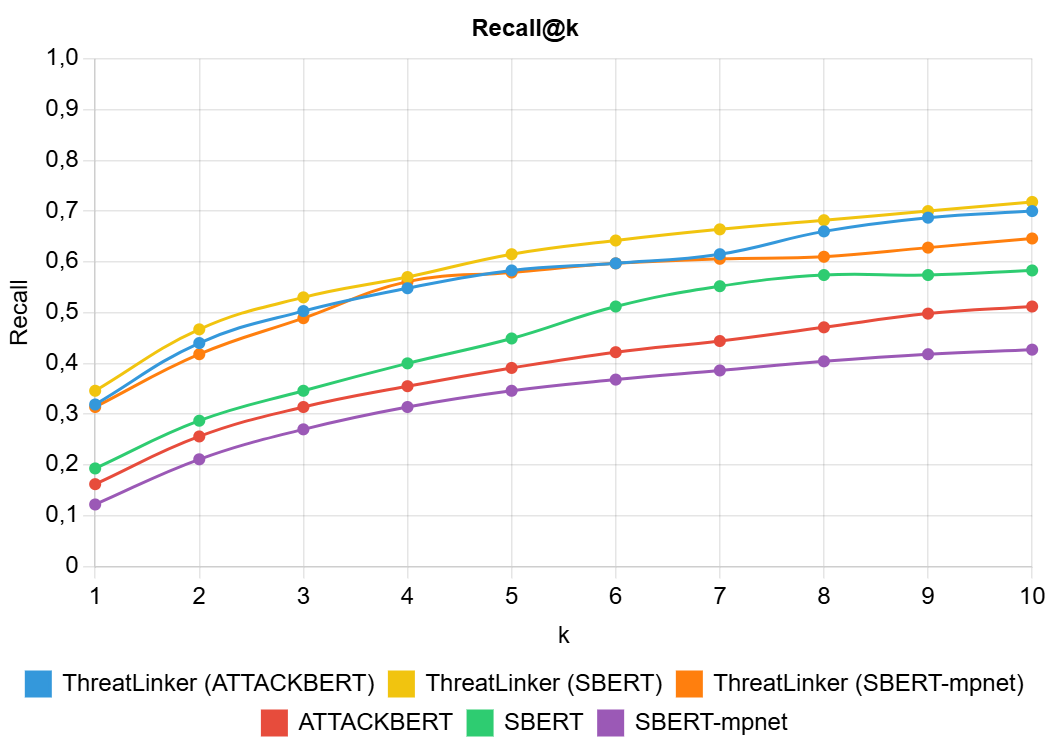}}}
    \qquad
    \subfloat[Precision@K.\label{fig:precision-k2}]{{\includegraphics[width=0.45\linewidth]{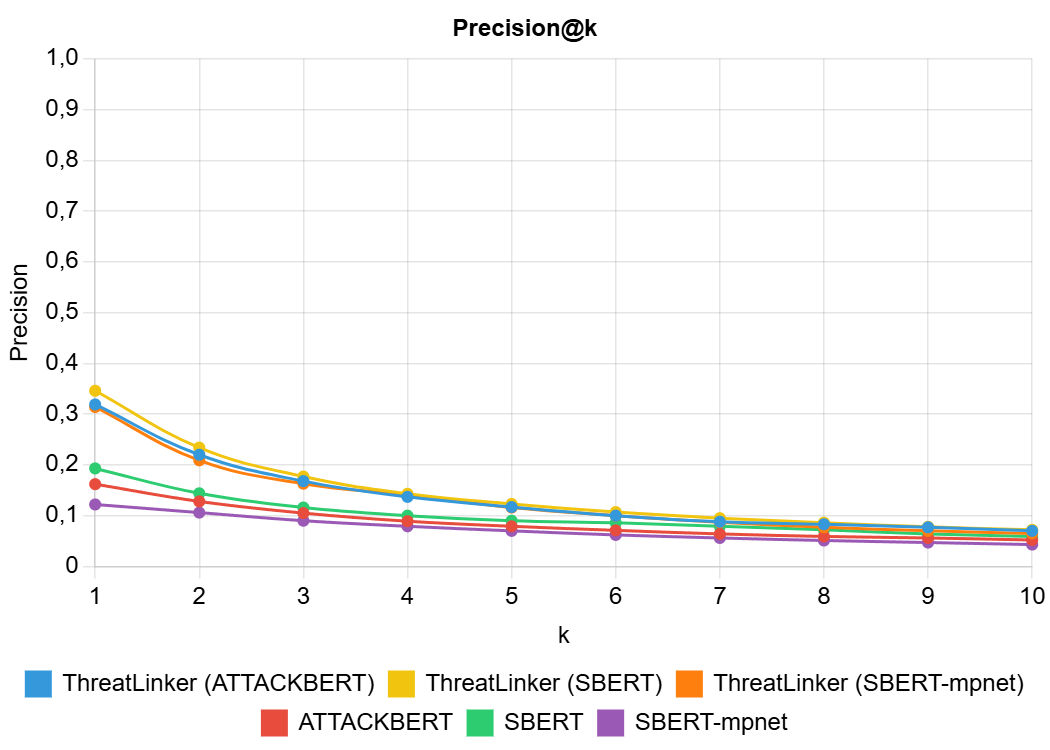}}}
    \caption{Recall@$K$ and Precision@$K$ for GT2.}
    \label{fig:recall_precision_k2}
\end{figure*}

The second ground truth (GT2) extends validation with $223$ CVE--CAPEC pairs across $160$ CVEs and $118$ CAPECs.
\Cref{fig:recall_precision_k2} shows that \oursystem (SBERT) achieves the highest recall ($>0.7$ for $K=10$) and precision across all $K$.
At $K=1$, only our models exceed $0.3$ recall, while state-of-the-art ones stay below $0.43$ even for $K=10$.
This analysis is confirmed by the Precision@K trend in \Cref{fig:precision-k2}.
\begin{wraptable}{r}{0.25\textwidth}
    %\vspace{-10pt}
    \centering
        \resizebox{0.25\textwidth}{!}{%
        \begin{tabular}{|c|c|}
            \hline
            \textbf{Model} & \textbf{MRR} \\ \hline
            \oursystem (ATT.BERT) & 0.546 \\ \hline
            \oursystem (SBERT) & \textbf{0.585} \\ \hline
            \oursystem (mpnet) & 0.539 \\ \hline
            SBERT & 0.397 \\ \hline
            ATTACKBERT & 0.357 \\ \hline
            SBERT (mpnet) & 0.294 \\ \hline
        \end{tabular}
        }
    \caption{MRR values (GT2).}
    \label{table:mrr_gt2}
    \vspace{-8pt}
\end{wraptable}
\Cref{table:mrr_gt2} reports MRR values, confirming superior ranking performance of \oursystem (SBERT, $0.585$) and \oursystem (ATTACKBERT, $0.546$), compared to the best baseline ($0.397$).
Results from GT2 reinforce GT1 findings: \oursystem consistently outperforms existing models across all metrics, with ATTACKBERT and SBERT ``paraphrase-mpnet-base-v2'' variants showing the most notable gains when combined with keyword search.

\review{}{While \oursystem consistently outperforms existing methods in terms of Recall@K and MRR on both GT1 and GT2, the absolute values of Precision@K remain relatively low.
For all configurations, Precision@K does not exceed $0.5$ on GT1 and $0.4$ on GT2, and the best MRR values ($0.587$ for GT1, $0.585$ for GT2) indicate that the first correct CAPEC association is often not ranked in the very top position. 
This behavior is partly explained by the nature of the task and the evaluation protocol. 
In fact, each CVE can plausibly be associated with multiple CAPEC patterns, but the ground truths only mark a small subset as relevant, so any additional plausible CAPECs in the top‑K are counted as false positives, which depresses precision.
On the other hand, many CAPEC entries are semantically close or belong to the same family (e.g., parent/child patterns), so the model may rank near‑synonymous or closely related patterns above the one selected in the ground truth, again harming Precision@K and MRR even when the suggested attack semantics are still meaningful for an analyst.}

\subsection{Comparison with supervised approaches}

To complete the evaluation, we compare \oursystem with the supervised approach by Giannakopoulos and Maliatsos~\cite{giannakopoulos2023usage}.
Their method uses a private dataset linking vulnerabilities to custom-defined threats for training and testing, addressing a slightly different task --mapping CVE descriptions to threats rather than CAPEC attack patterns.

Since their dataset and model are unavailable, we created a preprocessing step to enable comparison. 
With the support of three security experts, we mapped CAPEC attack patterns to the threats in~\cite{giannakopoulos2023usage}, using the same ground truth. 
We employ \oursystem (ATTACKBERT), which achieved the best overall performance.
\review{Given the task differences, we define $recall@k$ as the ratio of vulnerabilities correctly associated with a threat (\emph{TH-X}) within the top $K$ positions, i.e., $\text{Recall@k} = \frac{|V_k|}{|V|}$, and $precision@k$ as the fraction of True Positives in the top $K$ ranks, $\text{Precision@k} = \frac{|V_k|}{|V| \cdot k}$ where $|V_k|$ is the number of vulnerabilities correctly associated with \emph{TH-X} at position $\leq k$ and $|V|$ is the total number of vulnerabilities associated with \emph{TH-X}.}{
We use the same metrics defined in \Cref{sec:exp_settings}, where we consider the different threat descriptions (\emph{TH-X}) instead of attack patterns.
}

\begin{figure*}[!ht]
    \centering
    \subfloat[Precision.\label{fig:precision1}\centering]{{\includegraphics[width=0.29\linewidth]{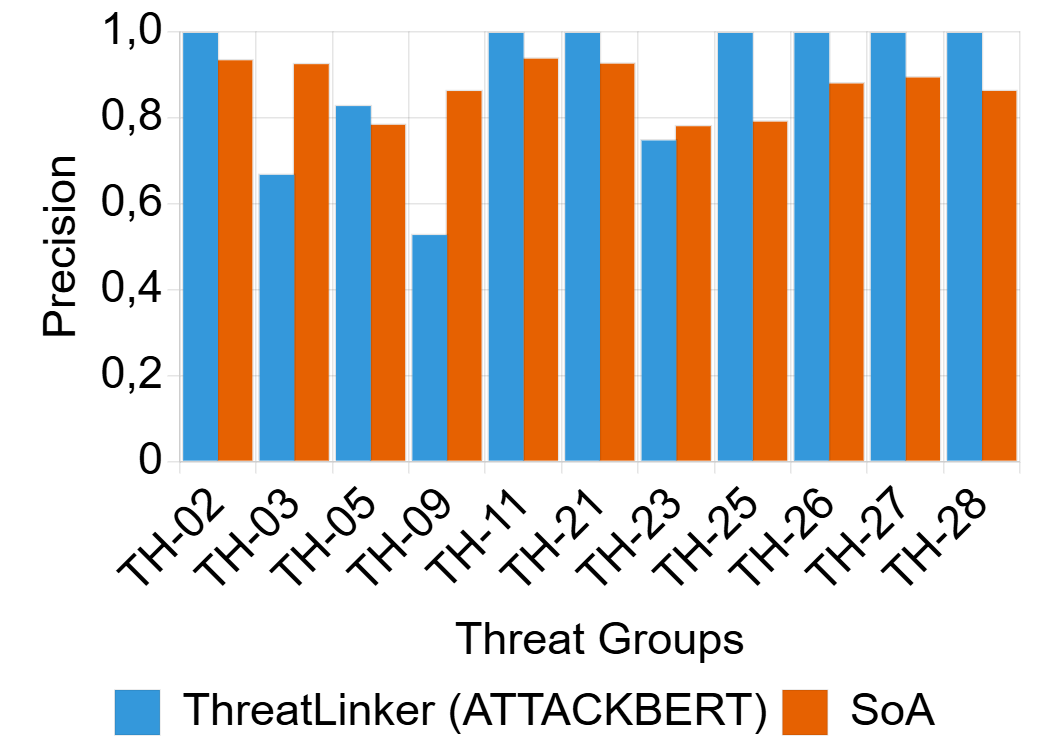}}}
    \qquad
    \subfloat[Recall.\label{fig:recall1}\centering]{{\includegraphics[width=0.29\linewidth]{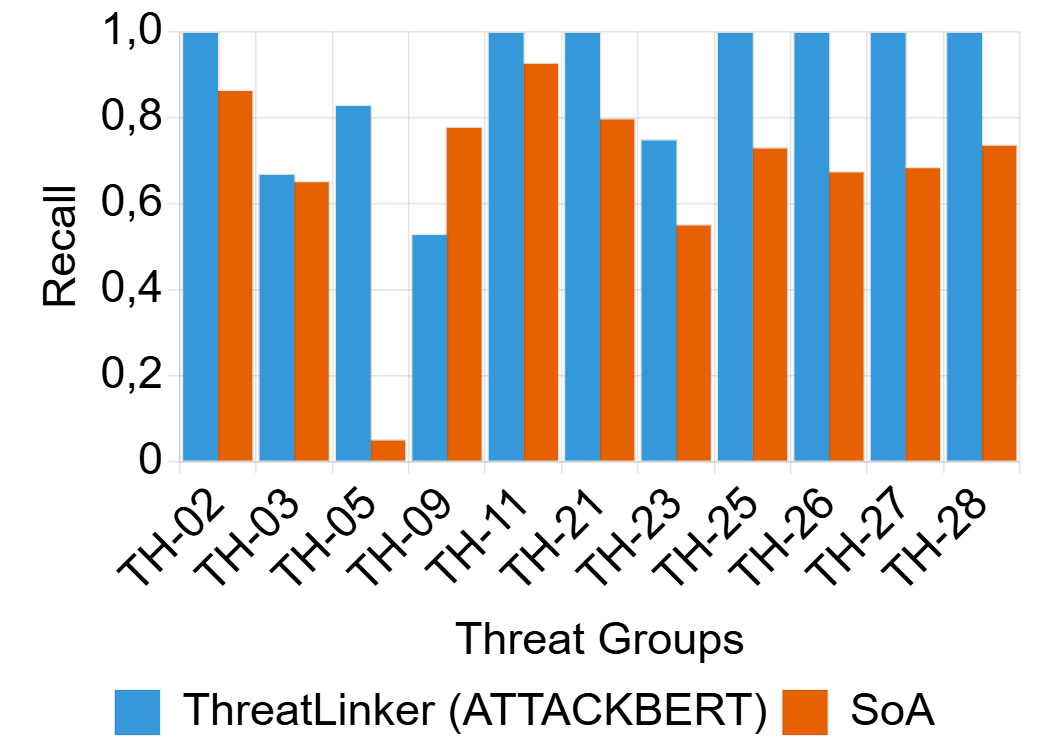}}}
    \qquad
    \subfloat[F1-score.\label{fig:f1score1}\centering]{{\includegraphics[width=0.29\linewidth]{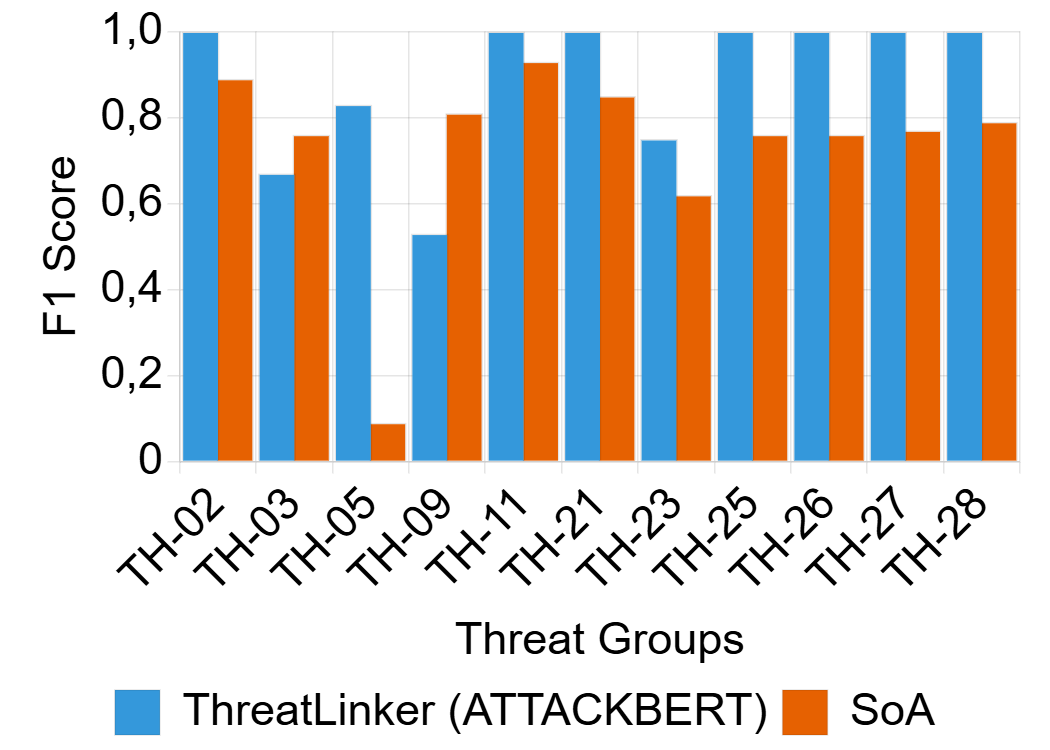}}}
    \qquad
    \caption{Comparison between \oursystem and the supervised approach.}
    \label{fig:recall_esorics}
\end{figure*}

\Cref{fig:recall_esorics} compares \oursystem (ATTACKBERT) with the state-of-the-art~\cite{giannakopoulos2023usage} for precision, recall, and F1 with $K{=}1$. 
For $K \geq 5$, \oursystem consistently identifies relevant threats, reaching recall values close to $1$. 
Even for Recall@1, it outperforms SoA in about $90\%$ of cases. 
Notably, our model excels for \textit{TH-05 (``Interception of Information'')} while shows only minor drops for the generic \textit{TH-09 (``Failure of System'')}, where CVE and CAPEC descriptions are less specific.

In summary, \oursystem demonstrates strong generalization to different tasks, such as mapping vulnerabilities to threats. 
It highlights the benefits of NLP-based methods, which require no training data, and the added value of keyword-based ranking in enhancing semantic similarity models.

\section{Threats to validity}
\label{sec:threatvalidity}

This paper introduced a novel methodology to associate attack patterns with vulnerabilities, supporting threat analysts in tasks such as risk assessment, intervention prioritization, and incident response.
By leveraging NLP, we significantly reduce the manual effort typically required for these analyses. 
Specifically, we proposed \oursystem, which combines keyword search and semantic similarity, advancing prior work relying solely on the latter~\cite{giannakopoulos2023usage,mdpi2022capeccve}.
We also discuss possible threats to validity to ensure the rigor of our findings.

\review{Our performance analysis shows that preprocessing vulnerability descriptions with regular expressions to remove version or build details enhances semantic similarity and overall accuracy.}{}
\oursystem performs best when vulnerability descriptions explicitly reference attack pattern attributes, even as acronyms. 
For instance, for CVE-1999-0906 (\emph{Buffer overflow in sccw allows local users to gain root access via the HOME environmental variable}), \oursystem correctly ranks CAPEC-10 (\emph{Buffer Overflow via Environment Variables}) as top-1, while ATTACKBERT and SBERT place it at rank 60.
Similar improvements occur for CVE-2006-6652 and CVE-2007-1057, where \oursystem ranks correct patterns at 3 and 2, compared to 26 and 32 for SoA models.

Some limitations arise when vulnerabilities mention multiple attack-related keywords, occasionally elevating irrelevant patterns. 
A potential solution is integrating context-aware NLP, e.g., syntactic or semantic parsing~\cite{nadeau2007survey}, or a multi-stage ranking using fine-tuned classifiers to distinguish direct from indirect associations. 
However, determining contextual relevance remains challenging, as crucial attack techniques may appear only once yet be conceptually central, highlighting current limits in model interpretability and semantic depth.

% Another insight is that in 19\% of errors, the misclassified patterns are parent or child CAPEC entries.
\review{Another insight is that often the misclassified patterns are parent or child CAPEC entries.}{Another current limitation is the relatively low Precision@K values.
Although the method consistently improves over existing baselines, correct CAPEC associations are not always placed at the very top of the ranked list. 
This is partly due to the evaluation protocol, which treats only a small set of CAPECs per CVE as relevant and penalizes all other candidates, even when they are semantically close or belong to the same CAPEC family. 
Consequently, the reported Precision@K values should be regarded as conservative.
To address this problem, a possible insight is the consideration of parent/child relations between attack patterns.
}
Incorporating this hierarchy allows control over attack pattern granularity, linking to parent patterns for generic CVEs and child patterns for detailed ones, thus preserving semantic coherence and improving ranking validity.
Nonetheless, associating a CVE with a CAPEC family rather than a single pattern may lead to overfitting. 
% Our implementation optionally supports this mode, though it is not part of the default methodology.
Integrating CAPEC hierarchies can enhance downstream security operations by supporting higher-level reasoning on related attacks, improving rule prioritization, and aligning with evolving TTPs.

\review{}{Further, the manually constructed dataset (GT2) introduces threats to validity regarding coverage and representativeness. 
The dataset is not a purely random sample of all CVEs, but after observing a strong over‑representation of XSS and SQL injection under fully random sampling, the selection was changed to a structured pseudo‑random strategy: random CVEs are then filtered to diversify CAPEC patterns. 
This improves heterogeneity but also biases the dataset towards CAPEC families that are easier to map from text and under-represent sparse, generic, or ambiguous cases. 
As a result, the dataset represents a curated benchmark to evaluate performance on a broad but limited subset of CVE--CAPEC relationships, rather than a statistically representative sample of the CVE and CAPEC ecosystems, which constrains the strength of claims about generalization to real‑world deployments.}
% 
% Finally, \oursystem could assist CVE Numbering Authorities (CNAs) in refining CVE descriptions. 
% We recommend enriching records with structured, semantically consistent details aligned with controlled vocabularies, an achievable improvement that would greatly benefit threat intelligence workflows.
\section{Conclusion}
\label{sec:conclusion}

\iffalse
This paper contributes \oursystem for associating attack patterns with vulnerabilities to support threat analysts in tasks such as risk assessment, intervention prioritization, and incident response. 
By leveraging NLP models and automation, we reduce the substantial manual effort typically required.
The performance is improved with respect to state-of-the-art techniques thanks to the combination of keyword search with semantic similarity.
% 
In future works, we will integrate the CAPEC relations within the model and examine how additional attributes (e.g., vector strings, CVSS metrics) can benefit the proposed solution.
% 
An alternative solution can involve machine learning with supervised training. A big challenge of this approach is the lack of datasets with attack pattern-vulnerability associations. Although we moved a step towards this problem by contributing a novel dataset with $223$ associations, we will work on developing a bigger one.
% 
Finally, to support long-term performance and continuous learning, a feedback mechanism could be established, allowing both human experts and automated systems (like \oursystem) to submit suggested CAPEC or CVE refinements. 
We believe \oursystem opens research opportunities as governing bodies like MITRE or CNAs could integrate and enhance cybersecurity catalogs for threat intelligence. 
This iterative approach would create a more accurate and richly annotated corpus of CVEs that reflects the real-world evolution of attack patterns.
\fi

This paper presents \oursystem, a framework that associates attack patterns with vulnerabilities to assist advanced threat analysis.
% in risk assessment, intervention prioritization, and incident response. 
By combining keyword search with semantic similarity, \oursystem improves performance over state-of-the-art methods while significantly reducing manual effort through NLP-based automation. 
Future work includes integrating CAPEC hierarchical relations and exploring additional attributes (e.g., vector strings, CVSS metrics) to enhance accuracy. 
We also plan to expand our current dataset of $223$ associations to enable supervised learning approaches, currently limited by the scarcity of labeled data. 
To ensure continuous improvement, a feedback mechanism could be established where experts and automated systems (like \oursystem) propose CAPEC or CVE refinements. 
Ultimately, \oursystem paves the way for collaboration with governing bodies such as MITRE and CVE Numbering Authorities to evolve cybersecurity catalogs and create richer, more accurate CVE annotations reflecting real-world attack dynamics.

\section*{Acknowledgements}
This work was partially supported by project SERICS (PE00000014) under the MUR National Recovery and Resilience Plan funded by the European Union - NextGenerationEU.

\bibliographystyle{IEEEtran}
\bibliography{mybibliography}

\end{document}